\documentclass[fleqn,twoside]{article}
\usepackage{espcrc2}
\usepackage{graphicx}

\newcommand{\AmS}{{\protect\the\textfont2
  A\kern-.1667em\lower.5ex\hbox{M}\kern-.125emS}}

\title{The Quest for Light Sea Quarks: Algorithms for the Future}

\author{W. Schroers\address[UWUP]{Department of Physics, University of
    Wuppertal, D-42097 Wuppertal, Germany} \thanks{Email:
    Wolfram.Schroers@Feldtheorie.de}, N.  Eicker\addressmark[UWUP], M.
  D'Elia\address[GENOVA]{Dipartimento di Fisica dell'Universit{\`a} di
    Genova and INFN, I-16146, Genova, Italy}, Ph.\ de
  Forcrand\address[ETHZ]{Inst.~f{\"u}r Theoretische Physik, ETH
    H{\"o}nggerberg, CH-8093 Z{\"u}rich, Switzerland}, C.
  Gebert\address[DESY]{Theory Division, DESY, Notkestr. 85, D-22603
    Hamburg, Germany}, Th. Lippert\addressmark[UWUP], I.
  Montvay\addressmark[DESY], \\ B.  Orth\addressmark[UWUP], M.
  Pepe\address[PARIS]{Laboratoire de Physique Th\'eorique Universit\'e
    de Paris-Sud, B\^atiment 210 F-91405 Orsay-Cedex}, K.
  Schilling\addressmark[UWUP]}

\begin{document}

\begin{abstract}
  As part of a systematic algorithm study, we present first results on
  a performance comparison between a multibosonic algorithm and the
  hybrid Monte Carlo algorithm as employed by the SESAM collaboration.
  The standard Wilson fermion action is used on $32\times 16^3$
  lattices at $\beta = 5.5$.
\end{abstract} 
\maketitle 
\section{Introduction}
\label{sec:introduction}
The past six years have witnessed a number of large simulations of
full QCD with two degenerate sea quarks and Wilson like actions. All
of them were based on the hybrid Monte Carlo algorithm (HMC) which has
been systematically improved over the years. These simulations,
however, albeit being milestones for lattice QCD in the
pre-Teracomputing age~\cite{Simulations}, suffer from severe
restrictions that prevent them to be truly realistic: we need to go to
{\it lighter}\/ quark masses and operate with {\it three}\/ flavors
with masses eventually approaching the physical mass.

This situation has been motivating research and development of
alternative algorithms of the multibosonic variety (MBA) that promise
in principle to overcome some of the restrictions of HMC mentioned.
It has been shown that MBA can indeed be geared to reach the
efficiency of HMC~\cite{Alexandrou:2000ii} and encouraging results
have been achieved with MBA for the treatment of finite density
QCD~\cite{Hands:2000ei}, supersymmetric field
theories~\cite{Campos:1999du} as well as the case of three dynamical
fermion flavors~\cite{Gebert:2001ar}.

In this contribution we wish to describe our benchmarking of MBA,
performed in the quest to achieve lighter quark masses than previously
possible in the SESAM and other full QCD simulations. For details
regarding the HMC algorithm used see ref.~\cite{Lippert:2001ha}.

\section{Gauging performance}
\label{sec:gauging-performance}
The efficiency of an algorithm is most suitably quoted as the
computational cost, $E_{\mbox{\tiny ind}}$, for producing a
statistically independent vacuum gauge field configuration within the
Markov process, expressed in units of the number of necessary
Dirac-matrix vector $M\cdot\phi$ multiplies:
 \begin{equation}
   \label{eq:efficiency}
   E_{\mbox{\tiny ind}} = 2 \tau_{\mbox{\tiny int}}E_{\mbox{\tiny
       traj}}\,,
\end{equation}
with $\tau_{\mbox{\tiny int}}$ being the integrated autocorrelation
time measured in units of update sequences called ``trajectories''
each having a cost of $E_{\mbox{\tiny traj}}$.
\begin{table}[hb]
  \begin{tabular}[c]{*{4}{c}}
    \hline
    \multicolumn{4}{l}{Volume: $\Omega=32\times 16^3$, $\beta=5.5$}
    \\ \hline
    $\mathbf{\kappa}$ & $\mathbf{a/fm}$ & $\mathbf{r_0/a}$ &
    $\mathbf{m_\pi/m_\rho}$ \\
    \hline
    $0.159$ & $0.141$ & $4.39(3)$ & $0.800(10)$ \\
    $0.160$ & $0.117$ & $4.89(3)$ & $0.670(14)$ \\
    \hline
  \end{tabular}
  \caption{Working points chosen.}
  \label{tab:working-points}
\end{table}
Being particularly interested in the quark mass dependence of the
algorithms, we chose for our analysis two different quark mass
settings, as given in table~\ref{tab:working-points}. Note that both
quark masses are from the regime of current full QCD simulations.

\section{Defining the MBA algorithm}
\label{sec:defin-mba-algor}
The parameter space of MBA algorithms is large {\it w.r.t.}\/ the HMC
situation: MBA optimization proceeds (a) by finding a not too high
order polynomial approximating the inverse of the fermion matrix at
the given working point, (b) by devising an efficient ``trajectory''
in form of an updating sequence for the degrees of freedom and (c) by
exploiting a method to correct for the remaining systematic error.
Today, all MBAs use a noisy correction step for (c). For a detailed
discussion see e.g.~ref.~\cite{deForcrand:1999nw,Schroers:2001ph}.

For the construction of appropriate polynomials we followed the
prescription proposed in ref.~\cite{Borrelli:1996re,deForcrand:1999sv}
which employs the GMRES algorithm to find a non-hermitian
approximation of the inverse Wilson matrix with ultraviolet-filtering
on a small sample of thermalized gauge field configurations. This was
done to minimize the polynomial order and hence the number of boson
fields. Quadratically optimized polynomials~\cite{Montvay:1998vh} need
about $2-3$ times more boson fields for similar acceptance rates.

The other key element for the optimization of an MBA is the choice of
``trajectory'', $T$, {\it i.e.}, the basic updating pattern. For that
matter, one has to devise an appropriate blend of gauge and boson
field updates. We have chosen the ``trajectory'' to be
\begin{equation}
  T = CO_g\left(O_bO_g\right)^5H_b\left(O_gO_b\right)^5O_g\; ,
\end{equation}
with $C$ being the final accept/reject correction step, $O_g$ ($O_b$)
a gauge (boson) field overrelaxation sweep and $H_b$ a boson field
global quasi-heat bath~\cite{deForcrand:1999je}.

The number of boson fields is tuned such as to ensure a good
acceptance rate in the ball-park of $60 - 70 \%$. As a result we
obtained the values quoted in table~\ref{tab:mba-fieldnum}.
\begin{table}[htb]
  \begin{tabular}[c]{l*{3}{c}}
    \hline
    $\mathbf{\kappa}$ & $\mathbf{n}$ & $\mathbf{P_{\mbox{\tiny
          acc}}}$ & $\mathbf{\Delta\beta}$ \\ \hline
    $0.159$ & $24$ & $60.3\%$ & $0.876$    \\
    $0.160$ & $42$ & $65.9\%$ & $1.763$    \\ \hline
  \end{tabular}
  \caption{Number of boson fields $n$ required for the MBA with the
    resulting acceptance rates at the two working points and the shift
    of the gauge coupling used for UV-filtering.}
  \label{tab:mba-fieldnum}
\end{table}

\section{Results}
\label{sec:results}
The determination of accurate autocorrelation times,
$\tau_{\mbox{\tiny int}}$, needs very long Monte Carlo time histories.
The number of trajectories underlying the present analysis together
with our results for the efficiencies, $E_{\mbox{\tiny ind}}$, are
collected in table~\ref{tab:efficiencies}. At $\kappa = 0.160$, we
have also used the topological charge, $Q$, as a monitor for
autocorrelations.
\begin{table}[tbh]
  \begin{tabular}[c]{clcc}
    \hline
    $\mathbf{\kappa}$ & \textbf{Observ.} & \textbf{HMC} &
    \textbf{MBA} \\ \hline
    $0.159$ & Plaq.       & $0.55816(4)$ & $0.55804(7)$ \\
            & $(am_\pi)$  & $0.4406(33)$ & $0.4488(37)$ \\
            & $(am_\rho)$ & $0.5507(59)$ & $0.5635(83)$ \\
    \hline
    $0.160$ & Plaq.       & $0.56077(6)$ & $0.56067(5)$ \\
            & $(am_\pi)$  & $0.3041(36)$ & $0.3184(68)$ \\
            & $(am_\rho)$ & $0.4542(77)$ & $0.4674(81)$ \\
    \hline
  \end{tabular}
  \caption{Average plaquette, non-singlet pseudo-scalar $(am_\pi)$ and
    vector $(am_\rho)$ meson masses.}
  \label{tab:observables}
\end{table}
\begin{table*}[htb]
  \begin{tabular}[c]{cl*{6}{c}}
    \hline
    $\mathbf{\kappa}$ & \textbf{Alg.}
    & \textbf{\# trajectories}
    & $\mathbf{\tau_{\mbox{\tiny int}}(\mbox{plaq})}$
    & $\mathbf{E_{\mbox{\tiny ind}}(\mbox{plaq})}$
    & $\mathbf{E_{\mbox{\tiny ind}}(m_\pi)}$
    & $\mathbf{E_{\mbox{\tiny ind}}(m_\rho)}$
    & $\mathbf{E_{\mbox{\tiny ind}}(Q)}$              \\ \hline
    $0.159$ & HMC & $3521$ & $11.0(0.4)$ & $712(26)10^3$ 
                  & $<1620\; 10^3$ & $<1620\; 10^3$ & \\
            & MBA & $6217$ & $44.7(3.4)$ & $786(60)10^3$ 
                  & $528\; 10^3$   & $704 \; 10^3$  & \\
    \hline
    $0.160$ & HMC & $5003$ & $34.1(3.1)$ & $572(52)10^4$ 
                  & $350\: 10^4$  & $894\; 10^4$ & $850(34)\; 10^4$ \\
            & MBA & $9910$ & $61.1(4.1)$ & $216(14)10^4$
                  & $102\; 10^4$  & $70\; 10^4$  & $324(22)\; 10^4$ \\
    \hline
  \end{tabular}
  \caption{Lengths of MC runs, estimated plaquette autocorrelation
    times and total cost, $E_{\mbox{\tiny ind}}$, for producing {\it
      one}\/ configuration decorrelated {\it w.r.t.}\/ various
    observables, from HMC and MBA\@.}
  \label{tab:efficiencies}
\end{table*}

As a check for our simulation code we have computed the average
plaquette and (non-singlet) pseudo-scalar and vector meson mass
values, as shown in table~\ref{tab:observables}. They agree within
errors.

$E_{\mbox{\tiny ind}}(\mbox{plaq})$ has been obtained from direct
integration of the plaquette autocorrelation functions while for the
masses the jackknife blocking technique has been applied. The latter
method only allows for a rather crude estimate of autocorrelation
times, so that we concentrate mainly on the plaquette in the
following, where we have safe control over statistical errors:

The statistics at the point $\kappa=0.159$ covers about $320$
autocorrelation times for the HMC and $140$ for the MBA\@. At the
working point $\kappa=0.160$, the HMC series has a length of $140$ and
the MBA of $160$ times $\tau_{\mbox{\tiny int}}$. Hence, the
statistics is sufficient for the results to be conclusive.

From table~\ref{tab:efficiencies} one finds that $E_{\mbox{\tiny
    ind}}(\mbox{plaq})$ is about equal for MBA and HMC at the larger
quark mass. The total effort in the case of the HMC increases by a
factor of $8$ when stepping down in quark mass, while at the same time
the cost for MBA `only' rises by a factor of $2.7$. Consequently, the
MBA has become almost a factor of three more efficient than HMC.

\section{Conclusions} 
\label{sec:conclusions}
We have demonstrated in a realistic setting that MBA-type sampling
techniques are not at all inferior to the state-of-the-art HMC
techniques and appear to show superior scaling behavior in quark
masses. We have established here a better scaling behavior of MBA,
leading to a gain factor of almost three in favor of the particular
MBA variant at currently reachable quark masses.

It remains to be seen to what extent this encouraging result will
pertain in the regime of smaller quark masses. The ultimate aim is to
push full QCD simulation with Wilson like fermions below the point
$m_{\pi}/m_{\rho} \leq 0.5$.

Another point to check is whether the MBA variant considered in this
paper with its non-hermitian approximation will remain operational for
practical simulations in the deeper chiral regime. Because the GMRES
method suffers from numerical instabilities for increasing orders $n$,
one might have to take recourse --- even on $32\times 16^3$ lattices
--- to $128$-bit precision. A more extended MBA study in this
direction is in preparation~\cite{Schroers:2001pr}.

\section*{Acknowledgments}
N.E.~is supported under DFG grant Li701/3-1, M.P.~by the European
Community's Human potential program under HPRN-CT-2000-00145
Hadrons/Lattice QCD\@, and W.S.~by the DFG Graduiertenkolleg
``Feldtheoretische Methoden in der Elemen\-tar\-teil\-chen\-theorie
und Statistischen Physik''. The numerical productions were run on the
APE-100 systems installed at NIC Zeuthen.

\end{document}